\documentclass[aps,prapplied, twocolumn, reprint,superscriptaddress,showpac,longbibliography]{revtex4-1}

\usepackage[colorlinks=true,citecolor=blue,linkcolor=magenta]{hyperref}
\hypersetup{
pdfauthor = {},
colorlinks = true, linkcolor = blue, urlcolor=blue, bookmarksnumbered =  true}

\usepackage{units}
\usepackage{graphicx}
\usepackage{amsmath}
\usepackage{physics}
\usepackage{ulem}

\usepackage{graphicx}
\usepackage{dcolumn}
\usepackage{bm}

\usepackage[utf8]{inputenc}
\usepackage[T1]{fontenc}
\usepackage{mathptmx}

\begin{document}

\title{(111)-oriented, single crystal diamond tips for nanoscale scanning probe imaging of out-of-plane magnetic fields}
\author{D. Rohner}
\affiliation{Department of Physics, University of Basel, Klingelbergstrasse 82, Basel CH-4056, Switzerland}
\author{J. Happacher}
\affiliation{Department of Physics, University of Basel, Klingelbergstrasse 82, Basel CH-4056, Switzerland}
\author{P. Reiser}
\affiliation{Department of Physics, University of Basel, Klingelbergstrasse 82, Basel CH-4056, Switzerland}
\author{M. A. Tschudin}
\affiliation{Department of Physics, University of Basel, Klingelbergstrasse 82, Basel CH-4056, Switzerland}
\author{A. Tallaire}
\affiliation{LSPM-CNRS, UPR 3407, Universit\'e Paris 13, Sorbonne Paris Cit\'e, 99 avenue Jean-Baptiste Cl\'ement, 93430 Villetaneuse, France}
\author{J. Achard}
\affiliation{LSPM-CNRS, UPR 3407, Universit\'e Paris 13, Sorbonne Paris Cit\'e, 99 avenue Jean-Baptiste Cl\'ement, 93430 Villetaneuse, France}
\author{B. J. Shields}
\affiliation{Department of Physics, University of Basel, Klingelbergstrasse 82, Basel CH-4056, Switzerland}
\author{P. Maletinsky}
\email[Corresponding author: ]{patrick.maletinsky@unibas.ch}
\affiliation{Department of Physics, University of Basel, Klingelbergstrasse 82, Basel CH-4056, Switzerland}


\date{\today}

\begin{abstract}
We present an implementation of all-diamond scanning probes for scanning nitrogen-vacancy (NV) magnetometry fabricated from $(111)$-oriented diamond material. 
The realized scanning probe tips on average contain single NV spins, a quarter of which have their spin quantization axis aligned parallel to the tip direction. 
Such tips enable single-axis vector magnetic field imaging with nanoscale resolution, where the measurement axis is oriented normal to the scan plane. 
We discuss how these tips bring multiple practical advantages for NV magnetometry, in particular regarding quantitative analysis of the resulting data.
We further demonstrate the beneficial optical properties of NVs oriented along the tip direction, such as polarization-insensitive excitation, which simplifies optical setups needed for NV magnetometry. 
Our results will be impactful for scanning NV magnetometry in general and for applications in spintronics and the investigation of thin film magnets in particular. 
\end{abstract}

\maketitle

Scanning probe magnetometry using nitrogen-vacancy (NV) center electronic spins in diamond offers a unique combination of spatial resolution, magnetic field sensitivity and quantitative magnetic imaging\,\cite{Maletinsky_2012,Appel_2016}. These combined performance characteristics have led to room-temperature imaging of single electron spins\,\cite{Grinolds_2013}, nanoscale domains in multiferroics\,\cite{Gross_2017} and antiferromagnets\,\cite{Gross_2017,Appel_2019}, and to cryogenic experiments addressing superconductors\,\cite{Thiel_2016,Pelliccione_2016,Rohner_2018} and magnetism in atomically thin crystals\,\cite{Thiel_2019}. These and other studies have demonstrated how scanning NV magnetometers can yield valuable insights into materials of high scientific and technological interest, beyond the capacity of existing nanoscale imaging methods.

\begin{figure}[h!]
\includegraphics[width=8.6cm]{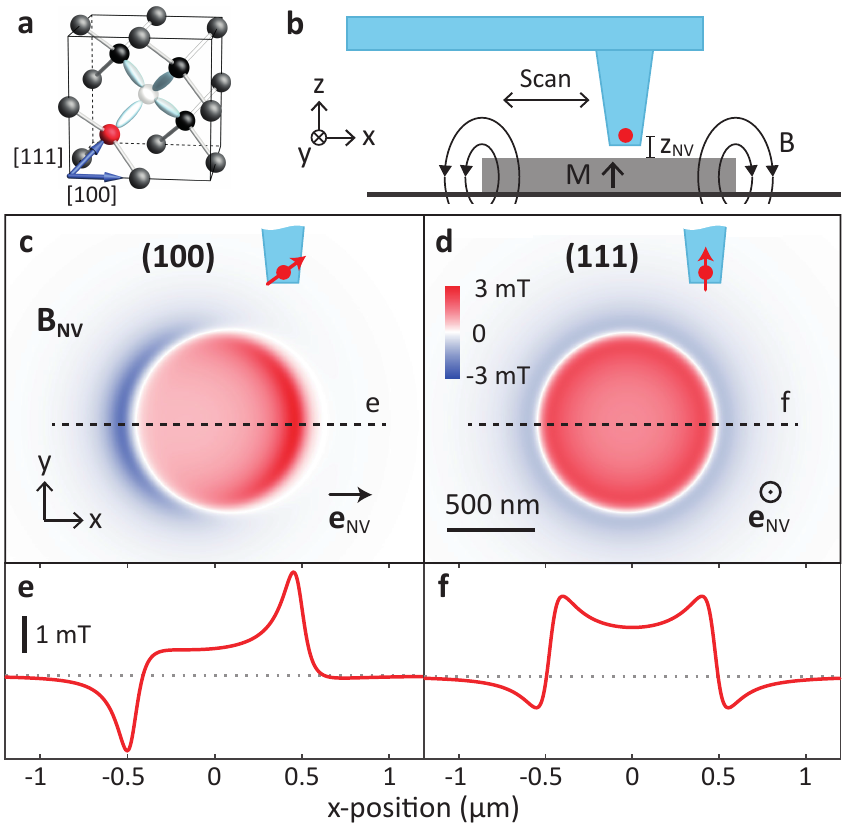}
\caption{\label{Schematic} (a) Atomic structure of the diamond NV center. Nitrogen atom and vacancy are represented by red and white spheres, with the $[111]$ and $[100]$ directions as indicated.
(b) Schematic of scanning nitrogen-vacancy (NV) center magnetometry with an all-diamond probe containing a single NV center (red point). The NV is scanned at height $z_{\rm NV}$ over a thin-film magnet (where typically, $z_{\rm NV}\sim30...80~$nm).
(c) Calculated magnetic field map $B_{\rm NV}(x,y)$ for a $(100)$-oriented diamond tip,  
with $z_{\rm NV}=70~$nm, disk diameter $1~\mu$m, and an areal disk magnetization $1~$mA.
(d) Same as in (c) but for an out-of-plane (OOP)-oriented NV in a $(111)$-oriented diamond tip. 
(e,f) Linecuts through the maps shown in (c,d).
Only for the OOP-oriented NV magnetometer, the rotational symmetry of the sample is retained in $B_{\rm NV}$.
}
\end{figure}

Magnetometry based on NV center spins\,\cite{Rondin_2014} builds on the extraordinary properties of this point defect in diamond that consists of a nitrogen atom adjacent to a lattice vacancy (Fig.\,\ref{Schematic}(a)). 
The negatively charged NV center has a ground state electronic spin-1, which is quantized along the NV binding axis, ${\bf e}_{\rm NV}$, and which can be readily initialized and read out by optical pumping and spin-dependent fluorescence\,\cite{Gruber_1997}.
These properties, along with the NV spin's response to magnetic fields through the Zeeman effect, render the NV spin an effective single-axis vector magnetometer, where, e.g., optically detected magnetic resonance (ODMR) can be employed for precise measurements of $B_{\rm NV}={\bf B}_{\rm ext}\cdot{\bf e}_{\rm NV}$, the projection of an external magnetic field ${\bf B}_{\rm ext}$ onto ${\bf e}_{\rm NV}$\,\cite{Rondin_2014}.

All-diamond scanning probes hosting individual NV spins in nanopillars (Fig.\,\ref{Schematic}(b)) offer a particularly powerful approach to perform nanoscale NV magnetometry\,\cite{Maletinsky_2012} and have been the basis of key achievements in the field\,\cite{Grinolds_2013,Gross_2017,Appel_2019,Appel_2015,Thiel_2016,Pelliccione_2016,Rohner_2018,Thiel_2019}.
All reported implementations of such scanning probes thus far consisted of diamond tips fabricated from $(100)$-oriented starting material -- the most commonly available crystal orientation for high purity diamonds. 
As a result, such $(100)$-oriented scanning probes yield nanoscale magnetic field maps of  $B_{\rm NV}$, where the measurement axis ${\bf e}_{\rm NV}\parallel \left<111\right>$ is oriented at $54.7^{\circ}$ from the sample normal. 
However, as will be discussed in the following, such vector magnetometry obtained with oblique measurement angles 
does not offer the optimal measurement configuration and could be significantly improved if NV centers oriented normal to the scanning plane could be employed.


It has been well established in scanning probe magnetic imaging that single-axis vector magnetometers measuring the out-of-plane (OOP) magnetic field component with respect to the scanning plane yield the best possible measurement configuration\,\cite{Lima_2009}, in particular to determine quantitative information about the underlying sample magnetization\,\cite{Casola_2018}. 
The key underlying reason is that a vector magnetic field image of any in-plane field component inherently lacks information about Fourier components of the magnetic field map corresponding to $k$-vectors perpendicular to the measurement axis. 
An OOP vector magnetometer is therefore the only configuration that avoids such loss of information, yielding a better signal-to-noise ratio in commonly employed reverse propagation methods\,\cite{Lima_2009,Roth1989}.
Additionally, stray field maps obtained with non-OOP magnetometers exhibit distortions, which complicate their direct interpretation, as illustrated by the calculated maps shown in Fig.\,\ref{Schematic}(c) and (d).
Besides these general advantages of an OOP measurement axis, the photophysics of the NV center demands that the NV axis  ${\bf e}_{\rm NV}$ be aligned to the external magnetic field for field amplitudes $\gtrsim10~$mT in order to maintain sizable ODMR contrasts\,\cite{Tetienne_2012}.
For studies of many prominent physical phenomena, such as quantum Hall\,\cite{Zhang_2005} or Skyrmion\,\cite{Nagaosa2013a} physics, this requirement can only be satisfied by an OOP NV orientation.
Lastly, an OOP NV orientation is also advantageous in that it maximizes the overlap of the NV optical dipoles with the waveguide mode of the scanning diamond nanopillar and thereby optimizes the optical addressing of NV spins\,\cite{Neu_2014}.


Based on this general motivation, we present an experimental realization of scanning NV magnetometry tips with OOP oriented NV spins. 
Such devices require the use of $(111)$-oriented diamond for scanning probe fabrication, where one of the four possible NV orientations is aligned with the $[111]$ axis and OOP with regard to the resulting scan plane.
Fabrication of diamond nanopillars on $(111)$-oriented diamond has been demonstrated previously\,\cite{Neu_2014}, however no scanning probe fabrication or NV magnetometry operation has been demonstrated for this crystal orientation thus far.
In this work we overcome this shortcoming, characterize the key properties of our resulting $(111)$-oriented, all-diamond scanning probes and experimentally demonstrate their performance in nanoscale magnetic imaging. 

\begin{figure}[t]
\includegraphics[width=8.6cm]{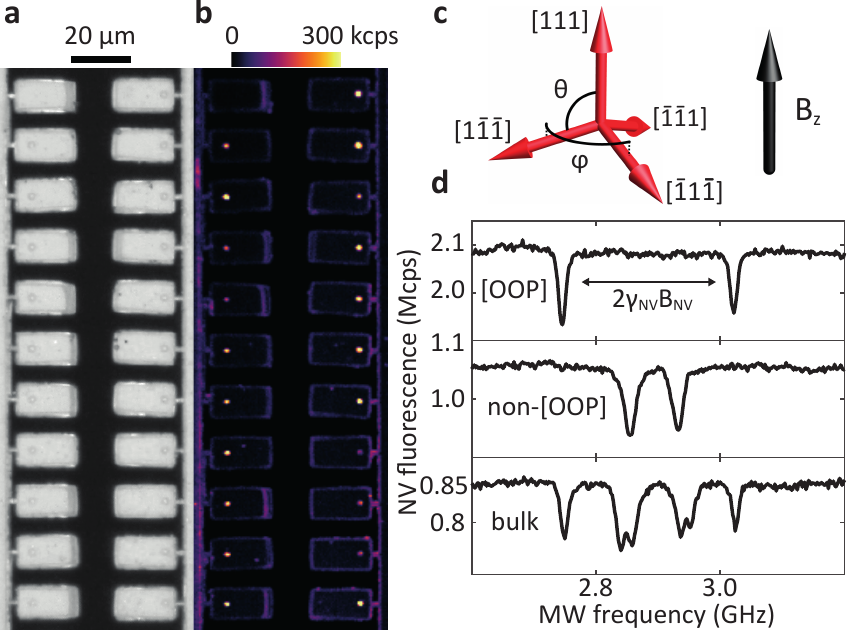}
\caption{\label{Characterization} (a) Optical microscopy picture of an array of finalized scanning probes fabricated from a $(111)$-oriented single crystal diamond. (b) Same area as in (a) imaged by confocal microscopy. Bright dots on the cantilevers correspond to nanopillars containing varying numbers of NV centers. (c) Visualization of the NV orientations and the applied magnetic field in $z$-direction. $\varphi$ and $\theta$ denote the angles in the spherical coordinate system. (d) ODMR spectra recorded on differently oriented NV centers in scanning probes and bulk diamond in an approximately OOP bias magnetic field of $5~$mT, yielding an almost perfect overlap of the three non-OOP NV orientations.}
\end{figure}

The starting point for our fabrication was electronic grade, $(111)$-oriented single crystal diamond (IIa technologies) which was laser-cut and polished to a $50~\mu$m thin diamond plate.
Near-surface NV centers were created by ion-implantation with $^{14}$N at $6~$keV and a density of $3 \times 10^{11}$~cm$^{-2}$, followed by vacuum annealing at $800~^\circ$C.
Following the fabrication recipe reported elsewhere\,\cite{Appel_2016}, this implantation density resulted in an average of two NV centers per scanning probe.
We obtained arrays of $10~\mu$m $\times~20~\mu$m sized cantilevers holding $\sim 300~$nm diameter nanopillars, where cantilevers and pillars had a depth of $2~\mu$m each (Fig.\,\ref{Characterization}(a)).
For scanning probe magnetometry, individual all-diamond cantilevers were detached from bulk diamond using micromanipulators and suitably attached to a quartz tuning fork for atomic force feedback control\,\cite{Appel_2016}. 

To characterize the resulting NV scanning probes for magnetometry, we performed confocal optical imaging (Fig.\,\ref{Characterization}(b)) using a homebuilt confocal microscope operating under ambient conditions.
The NV centers were addressed using a microscope objective with numerical aperture 0.8, excited non-resonantly with a green laser ($532~$nm) and the NV fluorescence was collected between $600~$nm and $800~$nm wavelength.
A metal wire placed close to the focal spot of the microscope was used to apply microwave magnetic fields to drive NV spins for ODMR.

Our fabrication resulted in a few hundred scanning probes on the diamond chip, a subset of which was investigated for the present study. 
Applying a magnetic field along the sample normal
allowed us to discriminate the OOP- (i.e. $[111]$)-oriented NVs from the non-OOP- (i.e. [1$\bar{1}\bar{1}$], [$\bar{1}$1$\bar{1}$] or [$\bar{1}\bar{1}$1])-oriented ones (Fig.\,\ref{Characterization}(c)).
Specifically, ODMR of OOP-oriented NVs shows a Zeeman splitting $2\gamma_{\rm NV} B_{\rm NV}$, with $\gamma_{\rm NV}=28~$GHz/T the NV's gyromagnetic ratio, that is three times larger than that of the other three orientations (Fig.\,\ref{Characterization}(d)). 


\begin{figure}[t]
\includegraphics[width=8.6cm]{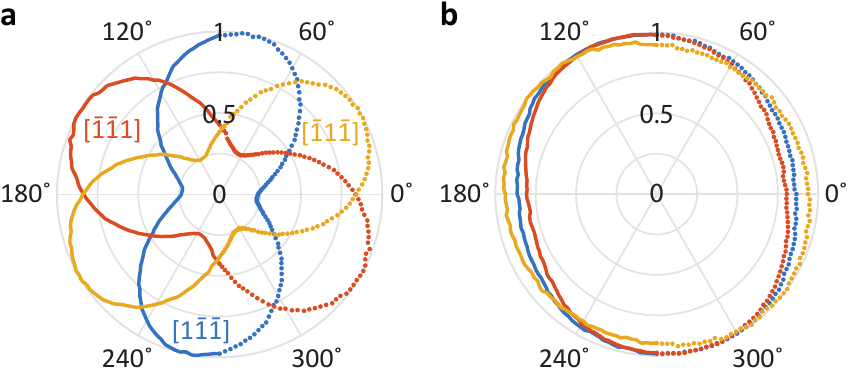}
\caption{\label{Polarization} (a) Dependence of NV fluorescence intensity on excitation polarization angle $\varphi_e$ for the three NV center orientations away from the out-of-plane (OOP) direction. The strong dependence on polarization angle allows for determination of the NV orientation. (b) Same as in (a) for three OOP-oriented NV centers, all of which show only a weak dependence on polarization angle. 
For both graphs, points show experimentally measured data, while lines correspond to the same data shifted by $180^\circ$.
For each dataset, fluorescence is normalized its maximal value.
}
\end{figure}

In addition, we explored the excitation polarization dependence of NV fluorescence in our devices. 
This allows for a microwave-free discrimination of NV orientations, based on the fact that the NV center's optical transitions are related to two optical dipoles lying in the plane orthogonal to the NV axis\,\cite{Maze_2011}.
For linearly polarized excitation well below saturation, the NV excitation rate is then directly proportional to the projection of the excitation polarization onto these dipoles. 
Basic geometrical considerations yield the dependence of fluorescence intensity $I_{\rm NV}$ on the angle of excitation polarization $\varphi_e$ 
as $I_{\rm NV}(\varphi_e) = I_{0}$ for OOP-oriented NVs and $I_{\rm NV}(\varphi_e) = I_{0} \left[\sin^2\left(\varphi_e - \varphi_{\rm NV}\right) + \frac{1}{9} \cos^2\left(\varphi_e - \varphi_{\rm NV}\right)\right]$ otherwise, where $\varphi_{\rm NV}$ is the azimuthal angle of the NV axis (cf. Fig.\,\ref{Characterization}(c))\,\cite{Alegre_2007, Lesik_2014}.


In our experiment, we used a liquid crystal polarization rotator (Thorlabs) to rotate the linear polarization of the excitation laser, set to a power below NV saturation, while collecting single NV fluorescence without polarization discrimination. 
Fig.\,\ref{Polarization}(a) shows data for three representative NVs oriented along the three non-OOP-oriented NV axes, where we observe distinct minima of $I(\varphi_e)$, reaching $\sim0.23~I_{0}$ at $\varphi_e = \varphi_{\rm NV}$ for each dataset. 
As expected, $\varphi_{\rm NV}$ is shifted by $\pm120^\circ$ between the three cases. The minima of $I(\varphi_e)$ exceed the expected value of $\frac{1}{9}I_{0}$ -- an observation we assign to background fluorescence and the onset of optical saturation.
For the OOP-oriented NV centers (Fig.\,\ref{Polarization}(b)), we observe a weak polarization dependence, which we attribute to mechanisms that break the cylindrical symmetry of the examined structures, such as off-center positioning of the NV centers in the pillar\,\cite{Bleuse_2011, Neu_2014}, asymmetric pillar shapes, or transverse strain experienced by the NV\,\cite{Barfuss_2019}.

\begin{figure}[t]
\includegraphics[width=8.6cm]{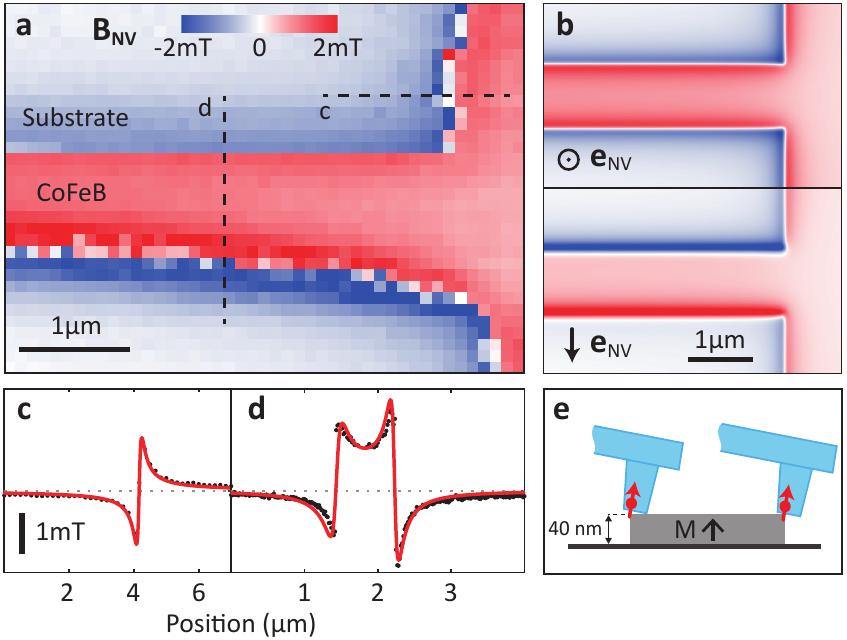}
\caption{\label{Magnetometry}
(a) Two-dimensional map of $B_{\rm NV}$ obtained with an OOP-oriented NV center on a patterned, $1~$nm thin CoFeB film in an OOP bias magnetic field of $4~$mT.
(b) Simulation of $B_{\rm NV}$ from a ``T-shaped'' sample as measured by an OOP NV (top) and by an NV with $\theta_\text{NV} = 54.7^\circ$ and $\varphi_\text{NV} = -90^\circ$ (bottom), using $M = 1~$mA and $z_{\rm NV} = 70~$nm. Same color scale as in (a).
(c)\&(d) Independently measured line cuts of $B_{\rm NV}$ along the lines indicated in (a). 
The red lines are analytical fits to the data (see text), which allow for a quantitative determination of $M$ and $z_{\rm NV}$. 
Importantly, and in contrast to other NV orientations, $B_{\rm NV}$ shows the same antisymmetric behavior across each sample edge, irrespective of edge orientation.
(e) Illustration of the tilt-induced variation of $z_{\rm NV}$ between adjacent edges, responsible for the unequal maxima of $B_{\rm NV}$ between the three CoFeB edges observed in (c)\&(d).
Note that the spatial sampling rate chosen for the image in (a) slightly exacerbates this asymmetry, compared to the more highly sampled line cuts in (c)\&(d).
}
\end{figure}

Finally, we conducted experiments to demonstrate the practical benefits of OOP-oriented single NV centers by using a corresponding tip for nanoscale magnetic imaging.
For this, we mapped stray magnetic fields emerging from a thin magnetic film (Ta/CoFeB($1~$nm)/MgO) with OOP magnetization, as commonly employed for experiments in spintronics and skyrmionics\,\cite{Tetienne_2014,Tetienne_2014b}. The sample was patterned into an arrangement of $1~\mu$m wide stripes to generate a non-trivial stray field distribution.
For magnetometry, the scanning NV center was positioned in the focus of the confocal microscope, while the sample was scanned laterally below the tip, with the distance between tip and sample stabilized by atomic force feedback\,\cite{Appel_2016}.
At each point of the scan, we recorded an ODMR spectrum to determine $B_{\rm NV}$.

A typical NV magnetometry image obtained over a ``T-shaped'' section of the CoFeB pattern is shown in Fig.\,\ref{Magnetometry}(a). 
The key outcome of employing an OOP-oriented NV for this experiment is that the measured stray field shows qualitatively the same behavior at every edge of the magnetic film, irrespective of edge orientation. 
Specifically, $B_{\rm NV}$ shows a sharp sign-reversal with a zero-crossing and a near-perfect antisymmetric behavior along the magnetic film edge (Fig.\,\ref{Magnetometry}(c)\&(d)).
This behavior is in strong contrast to magnetometry with non-OOP oriented scanning NV centers\,\cite{Thiel_2016,Rohner_2018,Appel_2019,Thiel_2019}, where the qualitative behavior of $B_{\rm NV}$ near sample edges depends strongly on the orientation of the edge (Fig.\,\ref{Magnetometry}(b)). 
The quantitative data obtained with our tip can be fitted by an analytical model, from which sample magnetization $M$ and NV-sample distance $z_{\rm NV}$ can be readily extracted\,\cite{Hingant_2015}. 
In the present case, where only the OOP component of the stray field is measured, the model for all edges takes the same and particularly simple form
\begin{equation}
B_{\rm NV}(\tilde{x})=B_z(\tilde{x})=-\frac{\mu_0 M}{2\pi}\frac{\tilde{x}}{\tilde{x}^2+z_{\rm NV}^2},
\end{equation}
where $\tilde{x}$ represents the lateral displacement from the magnet edge, $\mu_0$ the vacuum permeability and $M$ the two-dimensional sample magnetization. 
A fit to the linecuts in Fig.\,\ref{Magnetometry}(c)\&(d) gave $M=1.07\pm 0.02~$mA and $z_{\rm NV}=81\pm 17~$nm, $78\pm 2~$nm, and $55\pm 2~$nm for the three edges observed.
The difference in $z_{\rm NV}$ found between the different edges can be explained by geometrical factors originating from a small angular misalignment of the scanning tip with respect to the sample normal (Fig.\,\ref{Magnetometry}(e)). 
Importantly, we find in all cases that $z_{\rm NV}$ 
is clearly below $100~$nm, demonstrating the excellent spatial resolution of our $(111)$-oriented diamond probes in NV magnetometry.

In conclusion, we have demonstrated a realization of $(111)$-oriented diamond probes for scanning NV magnetometry, with NV spins oriented normal to the scan plane. With our experiments, including a nanoscale magnetic image using such tips, we have demonstrated how this measurement configuration brings practical advantages on various levels for NV magnetometry.
Next to the general advantages for scanning magnetometry, we have also demonstrated specific advantages with respect to the probe's nanophotonic properties. The insensitivity of the NV's optical response to the orientation of linearly polarized laser excitation reduces the required optical powers and lowers experimental complexity for an optimized sensing apparatus. As was shown in earlier works, the orientation of the optical dipoles of OOP-oriented NVs is beneficial also for NV fluorescence collection efficiency, and thereby for magnetometry sensitivity\,\citep{Alegre_2007, Neu_2014}.
Future improvements of our $(111)$-oriented diamond probes will further optimize this aspect and, together with the advantages for NV magnetometry demonstrated in this paper, will be beneficial for the wide range of applications of scanning NV magnetometry in nanomagnetism and mesoscopic physics. 

We thank V. Jacques, J.-V. Kim and J. K\"olbl for helpful comments and K. Garcia and R. Soucaille for fabrication of the CoFeB sample investigated. We gratefully acknowledge financial support through the NCCR QSIT (Grant No. 185902), the Basel QCQT doctoral school, through the EU Quantum Flagship project ASTERIQS (Grant No. 820394) and through the Swiss NSF Project Grants No. 169321 and 155845.


\begin{thebibliography}{27}%
\makeatletter
\providecommand \@ifxundefined [1]{%
 \@ifx{#1\undefined}
}%
\providecommand \@ifnum [1]{%
 \ifnum #1\expandafter \@firstoftwo
 \else \expandafter \@secondoftwo
 \fi
}%
\providecommand \@ifx [1]{%
 \ifx #1\expandafter \@firstoftwo
 \else \expandafter \@secondoftwo
 \fi
}%
\providecommand \natexlab [1]{#1}%
\providecommand \enquote  [1]{``#1''}%
\providecommand \bibnamefont  [1]{#1}%
\providecommand \bibfnamefont [1]{#1}%
\providecommand \citenamefont [1]{#1}%
\providecommand \href@noop [0]{\@secondoftwo}%
\providecommand \href [0]{\begingroup \@sanitize@url \@href}%
\providecommand \@href[1]{\@@startlink{#1}\@@href}%
\providecommand \@@href[1]{\endgroup#1\@@endlink}%
\providecommand \@sanitize@url [0]{\catcode `\\12\catcode `\$12\catcode
  `\&12\catcode `\#12\catcode `\^12\catcode `\_12\catcode `\%12\relax}%
\providecommand \@@startlink[1]{}%
\providecommand \@@endlink[0]{}%
\providecommand \url  [0]{\begingroup\@sanitize@url \@url }%
\providecommand \@url [1]{\endgroup\@href {#1}{\urlprefix }}%
\providecommand \urlprefix  [0]{URL }%
\providecommand \Eprint [0]{\href }%
\providecommand \doibase [0]{http://dx.doi.org/}%
\providecommand \selectlanguage [0]{\@gobble}%
\providecommand \bibinfo  [0]{\@secondoftwo}%
\providecommand \bibfield  [0]{\@secondoftwo}%
\providecommand \translation [1]{[#1]}%
\providecommand \BibitemOpen [0]{}%
\providecommand \bibitemStop [0]{}%
\providecommand \bibitemNoStop [0]{.\EOS\space}%
\providecommand \EOS [0]{\spacefactor3000\relax}%
\providecommand \BibitemShut  [1]{\csname bibitem#1\endcsname}%
\let\auto@bib@innerbib\@empty
\bibitem [{\citenamefont {Maletinsky}\ \emph {et~al.}(2012)\citenamefont
  {Maletinsky}, \citenamefont {Hong}, \citenamefont {Grinolds}, \citenamefont
  {Hausmann}, \citenamefont {Lukin}, \citenamefont {Walsworth}, \citenamefont
  {Loncar},\ and\ \citenamefont {Yacoby}}]{Maletinsky_2012}%
  \BibitemOpen
  \bibfield  {author} {\bibinfo {author} {\bibfnamefont {P.}~\bibnamefont
  {Maletinsky}}, \bibinfo {author} {\bibfnamefont {S.}~\bibnamefont {Hong}},
  \bibinfo {author} {\bibfnamefont {M.~S.}\ \bibnamefont {Grinolds}}, \bibinfo
  {author} {\bibfnamefont {B.}~\bibnamefont {Hausmann}}, \bibinfo {author}
  {\bibfnamefont {M.~D.}\ \bibnamefont {Lukin}}, \bibinfo {author}
  {\bibfnamefont {R.~L.}\ \bibnamefont {Walsworth}}, \bibinfo {author}
  {\bibfnamefont {M.}~\bibnamefont {Loncar}}, \ and\ \bibinfo {author}
  {\bibfnamefont {A.}~\bibnamefont {Yacoby}},\ }\bibfield  {title} {\enquote
  {\bibinfo {title} {A robust scanning diamond sensor for nanoscale imaging
  with single nitrogen-vacancy centres},}\ }\href {\doibase
  10.1038/nnano.2012.50} {\bibfield  {journal} {\bibinfo  {journal} {Nature
  Nanotechnology}\ }\textbf {\bibinfo {volume} {7}},\ \bibinfo {pages} {320}
  (\bibinfo {year} {2012})}\BibitemShut {NoStop}%
\bibitem [{\citenamefont {Appel}\ \emph {et~al.}(2016)\citenamefont {Appel},
  \citenamefont {Neu}, \citenamefont {Ganzhorn}, \citenamefont {Barfuss},
  \citenamefont {Batzer}, \citenamefont {Gratz}, \citenamefont {Tsch\"ope},\
  and\ \citenamefont {Maletinsky}}]{Appel_2016}%
  \BibitemOpen
  \bibfield  {author} {\bibinfo {author} {\bibfnamefont {P.}~\bibnamefont
  {Appel}}, \bibinfo {author} {\bibfnamefont {E.}~\bibnamefont {Neu}}, \bibinfo
  {author} {\bibfnamefont {M.}~\bibnamefont {Ganzhorn}}, \bibinfo {author}
  {\bibfnamefont {A.}~\bibnamefont {Barfuss}}, \bibinfo {author} {\bibfnamefont
  {M.}~\bibnamefont {Batzer}}, \bibinfo {author} {\bibfnamefont
  {M.}~\bibnamefont {Gratz}}, \bibinfo {author} {\bibfnamefont
  {A.}~\bibnamefont {Tsch\"ope}}, \ and\ \bibinfo {author} {\bibfnamefont
  {P.}~\bibnamefont {Maletinsky}},\ }\bibfield  {title} {\enquote {\bibinfo
  {title} {Fabrication of all diamond scanning probes for nanoscale
  magnetometry},}\ }\href {\doibase 10.1063/1.4952953} {\bibfield  {journal}
  {\bibinfo  {journal} {Review of Scientific Instruments}\ }\textbf {\bibinfo
  {volume} {87}},\ \bibinfo {pages} {063703} (\bibinfo {year}
  {2016})}\BibitemShut {NoStop}%
\bibitem [{\citenamefont {Grinolds}\ \emph {et~al.}(2013)\citenamefont
  {Grinolds}, \citenamefont {Hong}, \citenamefont {Maletinsky}, \citenamefont
  {Luan}, \citenamefont {Lukin}, \citenamefont {Walsworth},\ and\ \citenamefont
  {Yacoby}}]{Grinolds_2013}%
  \BibitemOpen
  \bibfield  {author} {\bibinfo {author} {\bibfnamefont {M.~S.}\ \bibnamefont
  {Grinolds}}, \bibinfo {author} {\bibfnamefont {S.}~\bibnamefont {Hong}},
  \bibinfo {author} {\bibfnamefont {P.}~\bibnamefont {Maletinsky}}, \bibinfo
  {author} {\bibfnamefont {L.}~\bibnamefont {Luan}}, \bibinfo {author}
  {\bibfnamefont {M.~D.}\ \bibnamefont {Lukin}}, \bibinfo {author}
  {\bibfnamefont {R.~L.}\ \bibnamefont {Walsworth}}, \ and\ \bibinfo {author}
  {\bibfnamefont {A.}~\bibnamefont {Yacoby}},\ }\bibfield  {title} {\enquote
  {\bibinfo {title} {Nanoscale magnetic imaging of a single electron spin under
  ambient conditions},}\ }\href {\doibase 10.1038/nphys2543} {\bibfield
  {journal} {\bibinfo  {journal} {Nature Physics}\ }\textbf {\bibinfo {volume}
  {9}},\ \bibinfo {pages} {215} (\bibinfo {year} {2013})}\BibitemShut {NoStop}%
\bibitem [{\citenamefont {Gross}\ \emph {et~al.}(2017)\citenamefont {Gross},
  \citenamefont {Akhtar}, \citenamefont {Garcia}, \citenamefont
  {Mart{\'{\i}}nez}, \citenamefont {Chouaieb}, \citenamefont {Garcia},
  \citenamefont {Carr{\'{e}}t{\'{e}}ro}, \citenamefont
  {Barth{\'{e}}l{\'{e}}my}, \citenamefont {Appel}, \citenamefont {Maletinsky},
  \citenamefont {Kim}, \citenamefont {Chauleau}, \citenamefont {Jaouen},
  \citenamefont {Viret}, \citenamefont {Bibes}, \citenamefont {Fusil},\ and\
  \citenamefont {Jacques}}]{Gross_2017}%
  \BibitemOpen
  \bibfield  {author} {\bibinfo {author} {\bibfnamefont {I.}~\bibnamefont
  {Gross}}, \bibinfo {author} {\bibfnamefont {W.}~\bibnamefont {Akhtar}},
  \bibinfo {author} {\bibfnamefont {V.}~\bibnamefont {Garcia}}, \bibinfo
  {author} {\bibfnamefont {L.~J.}\ \bibnamefont {Mart{\'{\i}}nez}}, \bibinfo
  {author} {\bibfnamefont {S.}~\bibnamefont {Chouaieb}}, \bibinfo {author}
  {\bibfnamefont {K.}~\bibnamefont {Garcia}}, \bibinfo {author} {\bibfnamefont
  {C.}~\bibnamefont {Carr{\'{e}}t{\'{e}}ro}}, \bibinfo {author} {\bibfnamefont
  {A.}~\bibnamefont {Barth{\'{e}}l{\'{e}}my}}, \bibinfo {author} {\bibfnamefont
  {P.}~\bibnamefont {Appel}}, \bibinfo {author} {\bibfnamefont
  {P.}~\bibnamefont {Maletinsky}}, \bibinfo {author} {\bibfnamefont {J.-V.}\
  \bibnamefont {Kim}}, \bibinfo {author} {\bibfnamefont {J.~Y.}\ \bibnamefont
  {Chauleau}}, \bibinfo {author} {\bibfnamefont {N.}~\bibnamefont {Jaouen}},
  \bibinfo {author} {\bibfnamefont {M.}~\bibnamefont {Viret}}, \bibinfo
  {author} {\bibfnamefont {M.}~\bibnamefont {Bibes}}, \bibinfo {author}
  {\bibfnamefont {S.}~\bibnamefont {Fusil}}, \ and\ \bibinfo {author}
  {\bibfnamefont {V.}~\bibnamefont {Jacques}},\ }\bibfield  {title} {\enquote
  {\bibinfo {title} {Real-space imaging of non-collinear antiferromagnetic
  order with a single-spin magnetometer},}\ }\href {\doibase
  10.1038/nature23656} {\bibfield  {journal} {\bibinfo  {journal} {Nature}\
  }\textbf {\bibinfo {volume} {549}},\ \bibinfo {pages} {252} (\bibinfo {year}
  {2017})}\BibitemShut {NoStop}%
\bibitem [{\citenamefont {Appel}\ \emph {et~al.}(2019)\citenamefont {Appel},
  \citenamefont {Shields}, \citenamefont {Kosub}, \citenamefont {Hedrich},
  \citenamefont {H\"ubner}, \citenamefont {Fa{\ss}bender}, \citenamefont
  {Makarov},\ and\ \citenamefont {Maletinsky}}]{Appel_2019}%
  \BibitemOpen
  \bibfield  {author} {\bibinfo {author} {\bibfnamefont {P.}~\bibnamefont
  {Appel}}, \bibinfo {author} {\bibfnamefont {B.~J.}\ \bibnamefont {Shields}},
  \bibinfo {author} {\bibfnamefont {T.}~\bibnamefont {Kosub}}, \bibinfo
  {author} {\bibfnamefont {N.}~\bibnamefont {Hedrich}}, \bibinfo {author}
  {\bibfnamefont {R.}~\bibnamefont {H\"ubner}}, \bibinfo {author}
  {\bibfnamefont {J.}~\bibnamefont {Fa{\ss}bender}}, \bibinfo {author}
  {\bibfnamefont {D.}~\bibnamefont {Makarov}}, \ and\ \bibinfo {author}
  {\bibfnamefont {P.}~\bibnamefont {Maletinsky}},\ }\bibfield  {title}
  {\enquote {\bibinfo {title} {Nanomagnetism of magnetoelectric granular
  thin-film antiferromagnets},}\ }\href {\doibase 10.1021/acs.nanolett.8b04681}
  {\bibfield  {journal} {\bibinfo  {journal} {Nano Letters}\ }\textbf {\bibinfo
  {volume} {19}},\ \bibinfo {pages} {1682} (\bibinfo {year}
  {2019})}\BibitemShut {NoStop}%
\bibitem [{\citenamefont {Thiel}\ \emph {et~al.}(2016)\citenamefont {Thiel},
  \citenamefont {Rohner}, \citenamefont {Ganzhorn}, \citenamefont {Appel},
  \citenamefont {Neu}, \citenamefont {M\"uller}, \citenamefont {Kleiner},
  \citenamefont {Koelle},\ and\ \citenamefont {Maletinsky}}]{Thiel_2016}%
  \BibitemOpen
  \bibfield  {author} {\bibinfo {author} {\bibfnamefont {L.}~\bibnamefont
  {Thiel}}, \bibinfo {author} {\bibfnamefont {D.}~\bibnamefont {Rohner}},
  \bibinfo {author} {\bibfnamefont {M.}~\bibnamefont {Ganzhorn}}, \bibinfo
  {author} {\bibfnamefont {P.}~\bibnamefont {Appel}}, \bibinfo {author}
  {\bibfnamefont {E.}~\bibnamefont {Neu}}, \bibinfo {author} {\bibfnamefont
  {B.}~\bibnamefont {M\"uller}}, \bibinfo {author} {\bibfnamefont
  {R.}~\bibnamefont {Kleiner}}, \bibinfo {author} {\bibfnamefont
  {D.}~\bibnamefont {Koelle}}, \ and\ \bibinfo {author} {\bibfnamefont
  {P.}~\bibnamefont {Maletinsky}},\ }\bibfield  {title} {\enquote {\bibinfo
  {title} {Quantitative nanoscale vortex imaging using a cryogenic quantum
  magnetometer},}\ }\href {\doibase 10.1038/nnano.2016.63} {\bibfield
  {journal} {\bibinfo  {journal} {Nature Nanotechnology}\ }\textbf {\bibinfo
  {volume} {11}},\ \bibinfo {pages} {677} (\bibinfo {year} {2016})}\BibitemShut
  {NoStop}%
\bibitem [{\citenamefont {Pelliccione}\ \emph {et~al.}(2016)\citenamefont
  {Pelliccione}, \citenamefont {Jenkins}, \citenamefont {Ovartchaiyapong},
  \citenamefont {Reetz}, \citenamefont {Emmanouilidou}, \citenamefont {Ni},\
  and\ \citenamefont {Jayich}}]{Pelliccione_2016}%
  \BibitemOpen
  \bibfield  {author} {\bibinfo {author} {\bibfnamefont {M.}~\bibnamefont
  {Pelliccione}}, \bibinfo {author} {\bibfnamefont {A.}~\bibnamefont
  {Jenkins}}, \bibinfo {author} {\bibfnamefont {P.}~\bibnamefont
  {Ovartchaiyapong}}, \bibinfo {author} {\bibfnamefont {C.}~\bibnamefont
  {Reetz}}, \bibinfo {author} {\bibfnamefont {E.}~\bibnamefont
  {Emmanouilidou}}, \bibinfo {author} {\bibfnamefont {N.}~\bibnamefont {Ni}}, \
  and\ \bibinfo {author} {\bibfnamefont {A.~C.~Bleszynski}\ \bibnamefont
  {Jayich}},\ }\bibfield  {title} {\enquote {\bibinfo {title} {Scanned probe
  imaging of nanoscale magnetism at cryogenic temperatures with a single-spin
  quantum sensor},}\ }\href {\doibase 10.1038/nnano.2016.68} {\bibfield
  {journal} {\bibinfo  {journal} {Nature Nanotechnology}\ }\textbf {\bibinfo
  {volume} {11}},\ \bibinfo {pages} {700} (\bibinfo {year} {2016})}\BibitemShut
  {NoStop}%
\bibitem [{\citenamefont {Rohner}\ \emph {et~al.}(2018)\citenamefont {Rohner},
  \citenamefont {Thiel}, \citenamefont {M\"uller}, \citenamefont {Kasperczyk},
  \citenamefont {Kleiner}, \citenamefont {Koelle},\ and\ \citenamefont
  {Maletinsky}}]{Rohner_2018}%
  \BibitemOpen
  \bibfield  {author} {\bibinfo {author} {\bibfnamefont {D.}~\bibnamefont
  {Rohner}}, \bibinfo {author} {\bibfnamefont {L.}~\bibnamefont {Thiel}},
  \bibinfo {author} {\bibfnamefont {B.}~\bibnamefont {M\"uller}}, \bibinfo
  {author} {\bibfnamefont {M.}~\bibnamefont {Kasperczyk}}, \bibinfo {author}
  {\bibfnamefont {R.}~\bibnamefont {Kleiner}}, \bibinfo {author} {\bibfnamefont
  {D.}~\bibnamefont {Koelle}}, \ and\ \bibinfo {author} {\bibfnamefont
  {P.}~\bibnamefont {Maletinsky}},\ }\bibfield  {title} {\enquote {\bibinfo
  {title} {Real-space probing of the local magnetic response of thin-film
  superconductors using single spin magnetometry},}\ }\href {\doibase
  10.3390/s18113790} {\bibfield  {journal} {\bibinfo  {journal} {Sensors}\
  }\textbf {\bibinfo {volume} {18}},\ \bibinfo {pages} {3790} (\bibinfo {year}
  {2018})}\BibitemShut {NoStop}%
\bibitem [{\citenamefont {Thiel}\ \emph {et~al.}(2019)\citenamefont {Thiel},
  \citenamefont {Wang}, \citenamefont {Tschudin}, \citenamefont {Rohner},
  \citenamefont {Guti{\'{e}}rrez-Lezama}, \citenamefont {Ubrig}, \citenamefont
  {Gibertini}, \citenamefont {Giannini}, \citenamefont {Morpurgo},\ and\
  \citenamefont {Maletinsky}}]{Thiel_2019}%
  \BibitemOpen
  \bibfield  {author} {\bibinfo {author} {\bibfnamefont {L.}~\bibnamefont
  {Thiel}}, \bibinfo {author} {\bibfnamefont {Z.}~\bibnamefont {Wang}},
  \bibinfo {author} {\bibfnamefont {M.~A.}\ \bibnamefont {Tschudin}}, \bibinfo
  {author} {\bibfnamefont {D.}~\bibnamefont {Rohner}}, \bibinfo {author}
  {\bibfnamefont {I.}~\bibnamefont {Guti{\'{e}}rrez-Lezama}}, \bibinfo {author}
  {\bibfnamefont {N.}~\bibnamefont {Ubrig}}, \bibinfo {author} {\bibfnamefont
  {M.}~\bibnamefont {Gibertini}}, \bibinfo {author} {\bibfnamefont
  {E.}~\bibnamefont {Giannini}}, \bibinfo {author} {\bibfnamefont {A.~F.}\
  \bibnamefont {Morpurgo}}, \ and\ \bibinfo {author} {\bibfnamefont
  {P.}~\bibnamefont {Maletinsky}},\ }\bibfield  {title} {\enquote {\bibinfo
  {title} {Probing magnetism in 2{D} materials at the nanoscale with
  single-spin microscopy},}\ }\href {\doibase 10.1126/science.aav6926}
  {\bibfield  {journal} {\bibinfo  {journal} {Science}\ }\textbf {\bibinfo
  {volume} {364}},\ \bibinfo {pages} {973--976} (\bibinfo {year}
  {2019})}\BibitemShut {NoStop}%
\bibitem [{\citenamefont {Rondin}\ \emph {et~al.}(2014)\citenamefont {Rondin},
  \citenamefont {Tetienne}, \citenamefont {Hingant}, \citenamefont {Roch},
  \citenamefont {Maletinsky},\ and\ \citenamefont {Jacques}}]{Rondin_2014}%
  \BibitemOpen
  \bibfield  {author} {\bibinfo {author} {\bibfnamefont {L.}~\bibnamefont
  {Rondin}}, \bibinfo {author} {\bibfnamefont {J.-P.}\ \bibnamefont
  {Tetienne}}, \bibinfo {author} {\bibfnamefont {T.}~\bibnamefont {Hingant}},
  \bibinfo {author} {\bibfnamefont {J.-F.}\ \bibnamefont {Roch}}, \bibinfo
  {author} {\bibfnamefont {P.}~\bibnamefont {Maletinsky}}, \ and\ \bibinfo
  {author} {\bibfnamefont {V.}~\bibnamefont {Jacques}},\ }\bibfield  {title}
  {\enquote {\bibinfo {title} {Magnetometry with nitrogen-vacancy defects in
  diamond},}\ }\href {http://dx.doi.org/10.1088/0034-4885/77/5/056503}
  {\bibfield  {journal} {\bibinfo  {journal} {Reports on Progress in Physics}\
  }\textbf {\bibinfo {volume} {77}},\ \bibinfo {pages} {56503} (\bibinfo {year}
  {2014})}\BibitemShut {NoStop}%
\bibitem [{\citenamefont {Gruber}\ \emph {et~al.}(1997)\citenamefont {Gruber},
  \citenamefont {Drabenstedt}, \citenamefont {Tietz}, \citenamefont {Fleury},
  \citenamefont {Wrachtrup},\ and\ \citenamefont
  {Borczyskowski}}]{Gruber_1997}%
  \BibitemOpen
  \bibfield  {author} {\bibinfo {author} {\bibfnamefont {A.}~\bibnamefont
  {Gruber}}, \bibinfo {author} {\bibfnamefont {A.}~\bibnamefont {Drabenstedt}},
  \bibinfo {author} {\bibfnamefont {C.}~\bibnamefont {Tietz}}, \bibinfo
  {author} {\bibfnamefont {L.}~\bibnamefont {Fleury}}, \bibinfo {author}
  {\bibfnamefont {J.}~\bibnamefont {Wrachtrup}}, \ and\ \bibinfo {author}
  {\bibfnamefont {C.}~\bibnamefont {Borczyskowski}},\ }\bibfield  {title}
  {\enquote {\bibinfo {title} {Scanning confocal optical microscopy and
  magnetic resonance on single defect centers},}\ }\href
  {http://dx.doi.org/10.1126/science.276.5321.2012} {\bibfield  {journal}
  {\bibinfo  {journal} {Science}\ }\textbf {\bibinfo {volume} {276}},\ \bibinfo
  {pages} {2012} (\bibinfo {year} {1997})}\BibitemShut {NoStop}%
\bibitem [{\citenamefont {Appel}\ \emph {et~al.}(2015)\citenamefont {Appel},
  \citenamefont {Ganzhorn}, \citenamefont {Neu},\ and\ \citenamefont
  {Maletinsky}}]{Appel_2015}%
  \BibitemOpen
  \bibfield  {author} {\bibinfo {author} {\bibfnamefont {P.}~\bibnamefont
  {Appel}}, \bibinfo {author} {\bibfnamefont {M.}~\bibnamefont {Ganzhorn}},
  \bibinfo {author} {\bibfnamefont {E.}~\bibnamefont {Neu}}, \ and\ \bibinfo
  {author} {\bibfnamefont {P.}~\bibnamefont {Maletinsky}},\ }\bibfield  {title}
  {\enquote {\bibinfo {title} {Nanoscale microwave imaging with a single
  electron spin in diamond},}\ }\href {\doibase 10.1088/1367-2630/17/11/112001}
  {\bibfield  {journal} {\bibinfo  {journal} {New Journal of Physics}\ }\textbf
  {\bibinfo {volume} {17}},\ \bibinfo {pages} {112001} (\bibinfo {year}
  {2015})}\BibitemShut {NoStop}%
\bibitem [{\citenamefont {Lima}\ and\ \citenamefont {Weiss}(2009)}]{Lima_2009}%
  \BibitemOpen
  \bibfield  {author} {\bibinfo {author} {\bibfnamefont {E.~A.}\ \bibnamefont
  {Lima}}\ and\ \bibinfo {author} {\bibfnamefont {B.~P.}\ \bibnamefont
  {Weiss}},\ }\bibfield  {title} {\enquote {\bibinfo {title} {Obtaining vector
  magnetic field maps from single-component measurements of geological
  samples},}\ }\href {\doibase 10.1029/2008jb006006} {\bibfield  {journal}
  {\bibinfo  {journal} {Journal of Geophysical Research}\ }\textbf {\bibinfo
  {volume} {114}},\ \bibinfo {pages} {B06102} (\bibinfo {year}
  {2009})}\BibitemShut {NoStop}%
\bibitem [{\citenamefont {Casola}\ \emph {et~al.}(2018)\citenamefont {Casola},
  \citenamefont {van~der Sar},\ and\ \citenamefont {Yacoby}}]{Casola_2018}%
  \BibitemOpen
  \bibfield  {author} {\bibinfo {author} {\bibfnamefont {F.}~\bibnamefont
  {Casola}}, \bibinfo {author} {\bibfnamefont {T.}~\bibnamefont {van~der Sar}},
  \ and\ \bibinfo {author} {\bibfnamefont {A.}~\bibnamefont {Yacoby}},\
  }\bibfield  {title} {\enquote {\bibinfo {title} {Probing condensed matter
  physics with magnetometry based on nitrogen-vacancy centres in diamond},}\
  }\href {http://dx.doi.org/10.1038/natrevmats.2017.88} {\bibfield  {journal}
  {\bibinfo  {journal} {Nature Reviews Materials}\ }\textbf {\bibinfo {volume}
  {3}},\ \bibinfo {pages} {17088} (\bibinfo {year} {2018})}\BibitemShut
  {NoStop}%
\bibitem [{\citenamefont {Roth}\ \emph {et~al.}(1989)\citenamefont {Roth},
  \citenamefont {Sepulveda},\ and\ \citenamefont {Wikswo}}]{Roth1989}%
  \BibitemOpen
  \bibfield  {author} {\bibinfo {author} {\bibfnamefont {B.~J.}\ \bibnamefont
  {Roth}}, \bibinfo {author} {\bibfnamefont {N.~G.}\ \bibnamefont {Sepulveda}},
  \ and\ \bibinfo {author} {\bibfnamefont {J.~P.}\ \bibnamefont {Wikswo}},\
  }\bibfield  {title} {\enquote {\bibinfo {title} {Using a magnetometer to
  image a two-dimensional current distribution},}\ }\href {\doibase
  10.1063/1.342549} {\bibfield  {journal} {\bibinfo  {journal} {J. Appl.
  Phys.}\ }\textbf {\bibinfo {volume} {65}},\ \bibinfo {pages} {361} (\bibinfo
  {year} {1989})}\BibitemShut {NoStop}%
\bibitem [{\citenamefont {Tetienne}\ \emph {et~al.}(2012)\citenamefont
  {Tetienne}, \citenamefont {Rondin}, \citenamefont {Spinicelli}, \citenamefont
  {Chipaux}, \citenamefont {Debuisschert}, \citenamefont {Roch},\ and\
  \citenamefont {Jacques}}]{Tetienne_2012}%
  \BibitemOpen
  \bibfield  {author} {\bibinfo {author} {\bibfnamefont {J.-P.}\ \bibnamefont
  {Tetienne}}, \bibinfo {author} {\bibfnamefont {L.}~\bibnamefont {Rondin}},
  \bibinfo {author} {\bibfnamefont {P.}~\bibnamefont {Spinicelli}}, \bibinfo
  {author} {\bibfnamefont {M.}~\bibnamefont {Chipaux}}, \bibinfo {author}
  {\bibfnamefont {T.}~\bibnamefont {Debuisschert}}, \bibinfo {author}
  {\bibfnamefont {J.-F.}\ \bibnamefont {Roch}}, \ and\ \bibinfo {author}
  {\bibfnamefont {V.}~\bibnamefont {Jacques}},\ }\bibfield  {title} {\enquote
  {\bibinfo {title} {Magnetic-field-dependent photodynamics of single {NV}
  defects in diamond: an application to qualitative all-optical magnetic
  imaging},}\ }\href {\doibase 10.1088/1367-2630/14/10/103033} {\bibfield
  {journal} {\bibinfo  {journal} {New Journal of Physics}\ }\textbf {\bibinfo
  {volume} {14}},\ \bibinfo {pages} {103033} (\bibinfo {year}
  {2012})}\BibitemShut {NoStop}%
\bibitem [{\citenamefont {Zhang}\ \emph {et~al.}(2005)\citenamefont {Zhang},
  \citenamefont {Tan}, \citenamefont {Stormer},\ and\ \citenamefont
  {Kim}}]{Zhang_2005}%
  \BibitemOpen
  \bibfield  {author} {\bibinfo {author} {\bibfnamefont {Y.}~\bibnamefont
  {Zhang}}, \bibinfo {author} {\bibfnamefont {Y.-W.}\ \bibnamefont {Tan}},
  \bibinfo {author} {\bibfnamefont {H.~L.}\ \bibnamefont {Stormer}}, \ and\
  \bibinfo {author} {\bibfnamefont {P.}~\bibnamefont {Kim}},\ }\bibfield
  {title} {\enquote {\bibinfo {title} {Experimental observation of the quantum
  {Hall} effect and {Berry's} phase in graphene},}\ }\href {\doibase
  10.1038/nature04235} {\bibfield  {journal} {\bibinfo  {journal} {Nature}\
  }\textbf {\bibinfo {volume} {438}},\ \bibinfo {pages} {201} (\bibinfo {year}
  {2005})}\BibitemShut {NoStop}%
\bibitem [{\citenamefont {Nagaosa}\ and\ \citenamefont
  {Tokura}(2013)}]{Nagaosa2013a}%
  \BibitemOpen
  \bibfield  {author} {\bibinfo {author} {\bibfnamefont {N.}~\bibnamefont
  {Nagaosa}}\ and\ \bibinfo {author} {\bibfnamefont {Y.}~\bibnamefont
  {Tokura}},\ }\bibfield  {title} {\enquote {\bibinfo {title} {Topological
  properties and dynamics of magnetic skyrmions},}\ }\href
  {https://doi.org/10.1038/nnano.2013.243} {\bibfield  {journal} {\bibinfo
  {journal} {Nature Nanotechnology}\ }\textbf {\bibinfo {volume} {8}},\
  \bibinfo {pages} {899} (\bibinfo {year} {2013})}\BibitemShut {NoStop}%
\bibitem [{\citenamefont {Neu}\ \emph {et~al.}(2014)\citenamefont {Neu},
  \citenamefont {Appel}, \citenamefont {Ganzhorn}, \citenamefont
  {Miguel-S{\'{a}}nchez}, \citenamefont {Lesik}, \citenamefont {Mille},
  \citenamefont {Jacques}, \citenamefont {Tallaire}, \citenamefont {Achard},\
  and\ \citenamefont {Maletinsky}}]{Neu_2014}%
  \BibitemOpen
  \bibfield  {author} {\bibinfo {author} {\bibfnamefont {E.}~\bibnamefont
  {Neu}}, \bibinfo {author} {\bibfnamefont {P.}~\bibnamefont {Appel}}, \bibinfo
  {author} {\bibfnamefont {M.}~\bibnamefont {Ganzhorn}}, \bibinfo {author}
  {\bibfnamefont {J.}~\bibnamefont {Miguel-S{\'{a}}nchez}}, \bibinfo {author}
  {\bibfnamefont {M.}~\bibnamefont {Lesik}}, \bibinfo {author} {\bibfnamefont
  {V.}~\bibnamefont {Mille}}, \bibinfo {author} {\bibfnamefont
  {V.}~\bibnamefont {Jacques}}, \bibinfo {author} {\bibfnamefont
  {A.}~\bibnamefont {Tallaire}}, \bibinfo {author} {\bibfnamefont
  {J.}~\bibnamefont {Achard}}, \ and\ \bibinfo {author} {\bibfnamefont
  {P.}~\bibnamefont {Maletinsky}},\ }\bibfield  {title} {\enquote {\bibinfo
  {title} {Photonic nano-structures on (111)-oriented diamond},}\ }\href
  {\doibase 10.1063/1.4871580} {\bibfield  {journal} {\bibinfo  {journal}
  {Applied Physics Letters}\ }\textbf {\bibinfo {volume} {104}},\ \bibinfo
  {pages} {153108} (\bibinfo {year} {2014})}\BibitemShut {NoStop}%
\bibitem [{\citenamefont {Maze}\ \emph {et~al.}(2011)\citenamefont {Maze},
  \citenamefont {Gali}, \citenamefont {Togan}, \citenamefont {Chu},
  \citenamefont {Trifonov}, \citenamefont {Kaxiras},\ and\ \citenamefont
  {Lukin}}]{Maze_2011}%
  \BibitemOpen
  \bibfield  {author} {\bibinfo {author} {\bibfnamefont {J.~R.}\ \bibnamefont
  {Maze}}, \bibinfo {author} {\bibfnamefont {A.}~\bibnamefont {Gali}}, \bibinfo
  {author} {\bibfnamefont {E.}~\bibnamefont {Togan}}, \bibinfo {author}
  {\bibfnamefont {Y.}~\bibnamefont {Chu}}, \bibinfo {author} {\bibfnamefont
  {A.}~\bibnamefont {Trifonov}}, \bibinfo {author} {\bibfnamefont
  {E.}~\bibnamefont {Kaxiras}}, \ and\ \bibinfo {author} {\bibfnamefont
  {M.~D.}\ \bibnamefont {Lukin}},\ }\bibfield  {title} {\enquote {\bibinfo
  {title} {Properties of nitrogen-vacancy centers in diamond: the group
  theoretic approach},}\ }\href {\doibase 10.1088/1367-2630/13/2/025025}
  {\bibfield  {journal} {\bibinfo  {journal} {New Journal of Physics}\ }\textbf
  {\bibinfo {volume} {13}},\ \bibinfo {pages} {025025} (\bibinfo {year}
  {2011})}\BibitemShut {NoStop}%
\bibitem [{\citenamefont {Alegre}\ \emph {et~al.}(2007)\citenamefont {Alegre},
  \citenamefont {Santori}, \citenamefont {Medeiros-Ribeiro},\ and\
  \citenamefont {Beausoleil}}]{Alegre_2007}%
  \BibitemOpen
  \bibfield  {author} {\bibinfo {author} {\bibfnamefont {T.~P.~Mayer}\
  \bibnamefont {Alegre}}, \bibinfo {author} {\bibfnamefont {C.}~\bibnamefont
  {Santori}}, \bibinfo {author} {\bibfnamefont {G.}~\bibnamefont
  {Medeiros-Ribeiro}}, \ and\ \bibinfo {author} {\bibfnamefont {R.~G.}\
  \bibnamefont {Beausoleil}},\ }\bibfield  {title} {\enquote {\bibinfo {title}
  {Polarization-selective excitation of nitrogen vacancy centers in diamond},}\
  }\href {\doibase 10.1103/physrevb.76.165205} {\bibfield  {journal} {\bibinfo
  {journal} {Physical Review B}\ }\textbf {\bibinfo {volume} {76}},\ \bibinfo
  {pages} {165205} (\bibinfo {year} {2007})}\BibitemShut {NoStop}%
\bibitem [{\citenamefont {Lesik}\ \emph {et~al.}(2014)\citenamefont {Lesik},
  \citenamefont {Tetienne}, \citenamefont {Tallaire}, \citenamefont {Achard},
  \citenamefont {Mille}, \citenamefont {Gicquel}, \citenamefont {Roch},\ and\
  \citenamefont {Jacques}}]{Lesik_2014}%
  \BibitemOpen
  \bibfield  {author} {\bibinfo {author} {\bibfnamefont {M.}~\bibnamefont
  {Lesik}}, \bibinfo {author} {\bibfnamefont {J.-P.}\ \bibnamefont {Tetienne}},
  \bibinfo {author} {\bibfnamefont {A.}~\bibnamefont {Tallaire}}, \bibinfo
  {author} {\bibfnamefont {J.}~\bibnamefont {Achard}}, \bibinfo {author}
  {\bibfnamefont {V.}~\bibnamefont {Mille}}, \bibinfo {author} {\bibfnamefont
  {A.}~\bibnamefont {Gicquel}}, \bibinfo {author} {\bibfnamefont {J.-F.}\
  \bibnamefont {Roch}}, \ and\ \bibinfo {author} {\bibfnamefont
  {V.}~\bibnamefont {Jacques}},\ }\bibfield  {title} {\enquote {\bibinfo
  {title} {Perfect preferential orientation of nitrogen-vacancy defects in a
  synthetic diamond sample},}\ }\href {\doibase 10.1063/1.4869103} {\bibfield
  {journal} {\bibinfo  {journal} {Applied Physics Letters}\ }\textbf {\bibinfo
  {volume} {104}},\ \bibinfo {pages} {113107} (\bibinfo {year}
  {2014})}\BibitemShut {NoStop}%
\bibitem [{\citenamefont {Bleuse}\ \emph {et~al.}(2011)\citenamefont {Bleuse},
  \citenamefont {Claudon}, \citenamefont {Creasey}, \citenamefont {Malik},
  \citenamefont {G{\'{e}}rard}, \citenamefont {Maksymov}, \citenamefont
  {Hugonin},\ and\ \citenamefont {Lalanne}}]{Bleuse_2011}%
  \BibitemOpen
  \bibfield  {author} {\bibinfo {author} {\bibfnamefont {J.}~\bibnamefont
  {Bleuse}}, \bibinfo {author} {\bibfnamefont {J.}~\bibnamefont {Claudon}},
  \bibinfo {author} {\bibfnamefont {M.}~\bibnamefont {Creasey}}, \bibinfo
  {author} {\bibfnamefont {N.~S.}\ \bibnamefont {Malik}}, \bibinfo {author}
  {\bibfnamefont {J.-M.}\ \bibnamefont {G{\'{e}}rard}}, \bibinfo {author}
  {\bibfnamefont {I.}~\bibnamefont {Maksymov}}, \bibinfo {author}
  {\bibfnamefont {J.-P.}\ \bibnamefont {Hugonin}}, \ and\ \bibinfo {author}
  {\bibfnamefont {P.}~\bibnamefont {Lalanne}},\ }\bibfield  {title} {\enquote
  {\bibinfo {title} {Inhibition, enhancement, and control of spontaneous
  emission in photonic nanowires},}\ }\href {\doibase
  10.1103/physrevlett.106.103601} {\bibfield  {journal} {\bibinfo  {journal}
  {Physical Review Letters}\ }\textbf {\bibinfo {volume} {106}},\ \bibinfo
  {pages} {103601} (\bibinfo {year} {2011})}\BibitemShut {NoStop}%
\bibitem [{\citenamefont {Barfuss}\ \emph {et~al.}(2019)\citenamefont
  {Barfuss}, \citenamefont {Kasperczyk}, \citenamefont {K\"olbl},\ and\
  \citenamefont {Maletinsky}}]{Barfuss_2019}%
  \BibitemOpen
  \bibfield  {author} {\bibinfo {author} {\bibfnamefont {A.}~\bibnamefont
  {Barfuss}}, \bibinfo {author} {\bibfnamefont {M.}~\bibnamefont {Kasperczyk}},
  \bibinfo {author} {\bibfnamefont {J.}~\bibnamefont {K\"olbl}}, \ and\
  \bibinfo {author} {\bibfnamefont {P.}~\bibnamefont {Maletinsky}},\ }\bibfield
   {title} {\enquote {\bibinfo {title} {Spin-stress and spin-strain coupling in
  diamond-based hybrid spin oscillator systems},}\ }\href
  {https://dx.doi.org/10.1103/PhysRevB.99.174102} {\bibfield  {journal}
  {\bibinfo  {journal} {Phys. Rev. B}\ }\textbf {\bibinfo {volume} {99}},\
  \bibinfo {pages} {174102} (\bibinfo {year} {2019})}\BibitemShut {NoStop}%
\bibitem [{\citenamefont {Tetienne}\ \emph {et~al.}(2014)\citenamefont
  {Tetienne}, \citenamefont {Hingant}, \citenamefont {Kim}, \citenamefont
  {Diez}, \citenamefont {Adam}, \citenamefont {Garcia}, \citenamefont {Roch},
  \citenamefont {Rohart}, \citenamefont {Thiaville}, \citenamefont
  {Ravelosona},\ and\ \citenamefont {Jacques}}]{Tetienne_2014}%
  \BibitemOpen
  \bibfield  {author} {\bibinfo {author} {\bibfnamefont {J.-P.}\ \bibnamefont
  {Tetienne}}, \bibinfo {author} {\bibfnamefont {T.}~\bibnamefont {Hingant}},
  \bibinfo {author} {\bibfnamefont {J.-V.}\ \bibnamefont {Kim}}, \bibinfo
  {author} {\bibfnamefont {L.~H.}\ \bibnamefont {Diez}}, \bibinfo {author}
  {\bibfnamefont {J.-P.}\ \bibnamefont {Adam}}, \bibinfo {author}
  {\bibfnamefont {K.}~\bibnamefont {Garcia}}, \bibinfo {author} {\bibfnamefont
  {J.-F.}\ \bibnamefont {Roch}}, \bibinfo {author} {\bibfnamefont
  {S.}~\bibnamefont {Rohart}}, \bibinfo {author} {\bibfnamefont
  {A.}~\bibnamefont {Thiaville}}, \bibinfo {author} {\bibfnamefont
  {D.}~\bibnamefont {Ravelosona}}, \ and\ \bibinfo {author} {\bibfnamefont
  {V.}~\bibnamefont {Jacques}},\ }\bibfield  {title} {\enquote {\bibinfo
  {title} {Nanoscale imaging and control of domain-wall hopping with a
  nitrogen-vacancy center microscope},}\ }\href {\doibase
  10.1126/science.1250113} {\bibfield  {journal} {\bibinfo  {journal}
  {Science}\ }\textbf {\bibinfo {volume} {344}},\ \bibinfo {pages} {1366}
  (\bibinfo {year} {2014})}\BibitemShut {NoStop}%
\bibitem [{\citenamefont {Tetienne}\ \emph {et~al.}(2015)\citenamefont
  {Tetienne}, \citenamefont {Hingant}, \citenamefont {Mart{\'\i}nez},
  \citenamefont {Rohart}, \citenamefont {Thiaville}, \citenamefont {Diez},
  \citenamefont {Garcia}, \citenamefont {Adam}, \citenamefont {Kim},
  \citenamefont {Roch}, \citenamefont {Miron}, \citenamefont {Gaudin},
  \citenamefont {Vila}, \citenamefont {Ocker}, \citenamefont {Ravelosona},\
  and\ \citenamefont {Jacques}}]{Tetienne_2014b}%
  \BibitemOpen
  \bibfield  {author} {\bibinfo {author} {\bibfnamefont {J.-P.}\ \bibnamefont
  {Tetienne}}, \bibinfo {author} {\bibfnamefont {T.}~\bibnamefont {Hingant}},
  \bibinfo {author} {\bibfnamefont {L.~J.}\ \bibnamefont {Mart{\'\i}nez}},
  \bibinfo {author} {\bibfnamefont {S.}~\bibnamefont {Rohart}}, \bibinfo
  {author} {\bibfnamefont {A.}~\bibnamefont {Thiaville}}, \bibinfo {author}
  {\bibfnamefont {L.~Herrera}\ \bibnamefont {Diez}}, \bibinfo {author}
  {\bibfnamefont {K.}~\bibnamefont {Garcia}}, \bibinfo {author} {\bibfnamefont
  {J.-P.}\ \bibnamefont {Adam}}, \bibinfo {author} {\bibfnamefont {J.-V.}\
  \bibnamefont {Kim}}, \bibinfo {author} {\bibfnamefont {J.-F.}\ \bibnamefont
  {Roch}}, \bibinfo {author} {\bibfnamefont {I.~M.}\ \bibnamefont {Miron}},
  \bibinfo {author} {\bibfnamefont {G.}~\bibnamefont {Gaudin}}, \bibinfo
  {author} {\bibfnamefont {L.}~\bibnamefont {Vila}}, \bibinfo {author}
  {\bibfnamefont {B.}~\bibnamefont {Ocker}}, \bibinfo {author} {\bibfnamefont
  {D.}~\bibnamefont {Ravelosona}}, \ and\ \bibinfo {author} {\bibfnamefont
  {V.}~\bibnamefont {Jacques}},\ }\bibfield  {title} {\enquote {\bibinfo
  {title} {{The nature of domain walls in ultrathin ferromagnets revealed by
  scanning nanomagnetometry}},}\ }\href {\doibase
  https://doi.org/10.1038/ncomms7733} {\bibfield  {journal} {\bibinfo
  {journal} {Nature Communications}\ }\textbf {\bibinfo {volume} {6}},\
  \bibinfo {pages} {6733} (\bibinfo {year} {2015})}\BibitemShut {NoStop}%
\bibitem [{\citenamefont {Hingant}\ \emph {et~al.}(2015)\citenamefont
  {Hingant}, \citenamefont {Tetienne}, \citenamefont {Mart\'{\i}nez},
  \citenamefont {Garcia}, \citenamefont {Ravelosona}, \citenamefont {Roch},\
  and\ \citenamefont {Jacques}}]{Hingant_2015}%
  \BibitemOpen
  \bibfield  {author} {\bibinfo {author} {\bibfnamefont {T.}~\bibnamefont
  {Hingant}}, \bibinfo {author} {\bibfnamefont {J.-P.}\ \bibnamefont
  {Tetienne}}, \bibinfo {author} {\bibfnamefont {L.~J.}\ \bibnamefont
  {Mart\'{\i}nez}}, \bibinfo {author} {\bibfnamefont {K.}~\bibnamefont
  {Garcia}}, \bibinfo {author} {\bibfnamefont {D.}~\bibnamefont {Ravelosona}},
  \bibinfo {author} {\bibfnamefont {J.-F.}\ \bibnamefont {Roch}}, \ and\
  \bibinfo {author} {\bibfnamefont {V.}~\bibnamefont {Jacques}},\ }\bibfield
  {title} {\enquote {\bibinfo {title} {Measuring the magnetic moment density in
  patterned ultrathin ferromagnets with submicrometer resolution},}\ }\href
  {https://dx.doi.org/10.1103/PhysRevApplied.4.014003} {\bibfield  {journal}
  {\bibinfo  {journal} {Phys. Rev. Applied}\ }\textbf {\bibinfo {volume} {4}},\
  \bibinfo {pages} {014003} (\bibinfo {year} {2015})}\BibitemShut {NoStop}%
\end{thebibliography}

%

\end{document}